\documentclass[11pt,prd,superscriptaddress,nofootinbib]{revtex4}
\usepackage{graphicx}

\usepackage[usenames,dvipsnames]{xcolor}

\usepackage{amssymb,amsmath,amsfonts}

\begin{document}
\begin{titlepage}
	
	\hfill\parbox{5cm} { }
	
	\vspace{25mm}
	
	\begin{center}
		{\Large \bf Temperature dependence of $\rho$ meson-nucleon coupling constant from the  AdS/QCD soft-wall model}
		
		\vskip 1. cm
		
		{Shahin Mamedov $^{a,b}$\footnote{ sh.mamedov62@gmail.com (corresponding author)}},
		{Narmin Nasibova $^{b}$\footnote{n.nesibli88@gmail.com}}
		\vskip 0.5cm
		
		{ \it \indent $^a$ Institute for Physical Problems, Baku State Universty, Z. Khalilov 23, Baku, AZ - 1148, Azerbaijan}\\
		\it \indent $^b$ Institute of Physics, Academy of Sciences,\\
		 H.Javid Avenue 131,  Baku AZ-1143, Azerbaijan\\
	\end{center}
	
	\thispagestyle{empty}
	
	\vskip2cm
	
	\centerline{\bf ABSTRACT}
	\vskip 4mm
	
	We investigate the dependence of the $\rho$ meson-nucleon coupling constant on the temperature of the medium using the soft-wall model of AdS/QCD.  The finite temperature profile functions for the vector and fermion fields are applied to the model having a thermal dilaton field. The interaction Lagrangian in the bulk between these fields is written as in zero temperature case and includes minimal- and magnetic-type interactions. The temperature dependence of the $g_{\rho NN}(T)$ coupling constant and its terms are plotted. We observe that the coupling constant and its separate terms become zero at the medium temperature near the Hawking temperature of phase transition.
	\vspace{2cm}
\end{titlepage}

\section{INTRODUCTION}

The study of hadron quantities such as mass, decay and coupling constants, form factors etc., at finite temperature is of great importance for the investigation of properties of hot hadronic matter obtained in the collisions of heavy ions or protons. After a correspondence was found in theory between five-dimensional gravity in anti-de Sitter space-time and quantum field theory in four dimensions \cite{1,2,3,4}, several models and approaches were constructed in order to solve the problems in particle physics and to investigate the processes, which take place in the hot hadronic matter and cannot be solved in perturbation theory. One such models is the so-called soft-wall model of the AdS/QCD correspondence \cite{5,6,7,8}. This model has turned out to be an effective tool for the calculation of phenomenological hadron quantities \cite{9,10,11,12,13,14,15,16,17,18,19}, for the phase transition between hot hadronic matter and  quark-gluon plasma \cite{20}, and for theoretical studies of hot hadronic matter \cite{21,22,23,24,25,26,27,28}. 
The properties of mesons and baryons in the thermal medium have been investigated in the framework of different models including the hard- and soft-wall models of the AdS/QCD bottom-up approach(see, for example, \cite{29} and references therein). The influence of the hot hadronic medium on interactions between hadrons is one of topical questions of elementary particle physics in the hot nuclear medium. In particular,  knowledge of the temperature dependence of the coupling constants and form factors of the strong interactions between mesons and baryons surrounded by hot matter will help us to understand the interactions between particles in this medium. AdS/QCD models are useful for study of this question as well. It is known that quantum field theory at 
finite temperature in a confined phase is holographically dual 
to gravitational theory in the AdS-Schwarzschild metric. The 
soft-wall AdS/QCD model at finite temperature is based on this gravity background. In addition to this thermalization, in
the thermal soft-wall model in \cite{23,25,26} the dilaton field,
which is thermal as well, was considered. This suggestion 
provides two sources for conformal symmetry breaking and 
an explicit form of such a dilaton was established using the 
thermal loop approach in a field theory \cite{25}. Solutions of the
equations of motion (profile functions) for the boson and 
fermion fields interacting with the thermal dilaton field were 
given in Refs. \cite{25,26}. As an application of this model,the 
hadron form factors and the electromagnetic properties of 
nucleons and Roper resonances at finite temperature were 
studied using this approach in Ref.\cite{27}. We are interested 
in the temperature dependence of the interactions between  the hadrons. More concretely, we want to know how much 
the screening constants of the meson-baryon couplings in 
hot hadronic matter will depend on temperature prior to 
the confinement-deconfinement phase transition. Having 
explicit profile functions for the hadrons and using the 
holographic technique, one can easily solve this problem in 
the framework of this thermal dilaton soft-wall model. Here 
we consider the simplest of these coupling constants— 
namely, the  $\rho$ meson-nucleon coupling constant in the framework of this approach- and investigate the dependence of this constant when the temperature approaches the  phase  transition  temperature.

 The remainder of this paper is organized as follows: In
 Sec. II we introduce the soft-wall model at finite temperature. In Secs. III and V we obtain the bulk-to-boundary propagators for the free vector, scalar, and spinor fields in the bulk at finite temperature. In Sec. IV we develop the idea of the chiral condensate and the breaking of chiral
 symmetry at finite temperature. In Sec. VI we write the
Lagrangian for the vector-spinor interaction in the bulk and,
 using holographic correspondence, obtain the temperature- dependent integral expression for the $\rho$ meson-nucleon coupling constant in boundary QCD theory. In Sec. VII we
 fix parameters and plot graphics for the dependence of the
 coupling constant on the temperature, and in Sec. VIII we
discuss  our  results.
 
\section{SOFT-WALL MODEL AT FINITE TEMPERATURE}
In general, in the soft-wall model of AdS/QCD at finite temperature the dilaton field  $\varphi(z)$ can be considered the temperature-dependent one, $\varphi(z,T)$ and the action for this model will be written in terms of such a dilaton:
\begin{equation}
S=\int d^{4}x dz\sqrt{g}e^{-\varphi(z,T)}L(x,z,T),
\label{1}
\end{equation}
Here $ g$ denotes $ g=|det g_{M N}|$ $(M,N=0,1,2,3,5)$ and the extra dimension $z$ varies in the range $0\leqslant  z< \infty$. An exponential factor was introduced to make the integral over the $z$ coordinate  finite at the IR boundary $(r\rightarrow \infty )$ and the $k$ parameter is a scale parameter of a few hundred $ MeV$.
According to the case with the AdS/CFT correspondence at finite temperature, the boundary field theory
 corresponds to the gravity theory given by the AdS-
 Schwarzschild  metric \cite{2}:
\begin{eqnarray}
ds^{2}=e^{2A(z)}\left[-f(z)dt^{2}-\left(d\vec{x}\right)^{2}-\frac{dz^2}{f(z)}\right],
\nonumber \\
 f(z)= 1-\frac{z^4}{z_H^4},
 \label{2}
\end{eqnarray}
where $z_{H}$ is the position of the event horizon and  is related to the Hawking temperature as $T=1/(\pi z_{H})$, $ x=(t,\vec x)$ is the set of Minkowski coordinates, $A(z)=log(\frac{R}{z})$, and  $R$ is the AdS radius. 

For convenience, in \cite{25,26} the Regge- Wheeler tortoise coordinate $r$,
\begin{equation*}
r=\int \frac{dz}{f(z)}
\end{equation*}
was applied as a fifth one instead of $z$, and  the  terms of higher order than $T^{8}$  in the expansion of $z$ were neglected. This gives following relation between the  $r$ and  $z$ coordinates:
\begin{equation}
r\approx z\left[1+\frac{z^{4}}{5z_{H}^{4}}+\frac{z^{8}}{9z_{H}^{8}}\right].
\label{3}
\end{equation}
The metric for the AdS-Schwarzschild space time in these coordinates will be written as 
\begin{equation}
ds^{2}=e^{2A(r)}f^{\frac{3}{5}}(r)\left[dt^{2}-\frac{\left(d\vec{x}\right)^{2}}{f(r)}-dr^{2}\right]
\label{4}
\end{equation}
with $A(r)=log(\frac{R}{r})$. The thermal factor $f(r)$ in terms of the $r$ coordinate has the same form as in Eq. (2):
\begin{equation}
f(r)=1-\frac{r^{4}}{r_{H}^{4}}.
\label{5}
\end{equation}

In the approach for the finite temperature soft-wall model in Refs. \cite{25,26,27} the dilaton field $\varphi=k^{2}z^{2}$ is considered  a temperature- dependent one when the temperature dependent  dilaton parameter $k^2$ is introduced: 
\begin{equation}
 \varphi(r,T)=K^{2}(T)r^{2},
 \label{6}
\end{equation}
This thermal form of the dilaton parameter has been established by means of the relation between $k^2$ and the quark condensate $\Sigma $. It was proposed the relation between these constants holds in a finite temperature case, and thus the temperature dependence of the condensate  $\Sigma (T) $, which is known from chiral perturbation theory,  determines the temperature dependence of the dilaton parameter  $K^{2}(T)$.  In such a way  $K^{2}(T)$ was finally found in the  following form \cite{25}:  

\begin{equation}
K^{2}(T)=k^{2}\left[1+\rho(T) \right].
\label{7}
 \end{equation}
 So, $K^{2}(T)$ is the parameter of spontaneous breaking of chiral symmetry and the thermal-dependent term  $\rho (T)$ up to $T^4$ order was established in the  form:
\begin{equation} 
\rho(T)=\delta_{T_{1}}\frac{T^{2}}{12F^{2}}+\delta_{T_{2}}\left(\frac{T^2}{12F^2}\right)^{2}.
\label{8}
\end{equation}
Here  $ \emph F$ is the pseudoscalar decay constant in the chiral limit and the coefficients $\delta_{T_{1}}$ and  $\delta_{T_{2}}$ are defined by the number of quark flavors $N_f$ as follows:
  
\begin{equation}
\delta_{T_{1}}=-\frac{ N_{f}^{2}-1}{N_{f}},
\label{9}
 \end{equation}
 and
\begin{equation}
\delta_{T_{2}}=-\frac{N_{f}^{2}-1}{2N_{f}^{2}}. 
\label{10} 
\end{equation}
 
\section{THE MESON PROFILE FUNCTION AT FINITE TEMPERATURE}
 Let us briefly present here the  derivation of the profile function for mesons in the model with the thermal dilaton, which was described in \cite{25} in detail. The vector field $M_{N}(x, r,T)$, which on the ultraviolet boundary of space-time ($r=0$) gives the wave function of the vector meson, is composed from the gauge fields $A_L$ and $A_R$: $M_{N}=1/2\left(A_L+A_R\right)$. These gauge fields transform under the flavor symmetry subgroups $SU(2)_L$ and $SU(2)_R$, respectively, which are part of the $SU(2)_L\times SU(2)_R$ flavor group of the model. From these chiral gauge fields can be  composed  an axial vector field as well, which we do not consider here. The action for the scalar and vector fields in the AdS-Schwarzschild space-time in the general reads as 
 \begin{eqnarray}
S_M=-\frac{1}{2}\int d^{4}xdr\sqrt{g}e^{-\varphi(r,T)}[\partial_{N}M_{N}(x, r,T)\partial^{N}M^{N}(x, r,T)\nonumber\\
-(\mu^{2}(r,T)+V(r,T))M_{N}(x, r,T)M^{N}(x, r,T)].\label{11}
\end{eqnarray}  
Here  $V(r,T)$  is the thermal dilaton potential, and it is expressed as 
\begin{equation}
V(r,T)=\frac{e^{-2A(r)}}{f^{\frac{3}{5}}(r)}\left[\varphi^{\prime\prime}(r,T)+\varphi^{\prime}(r,T){A}^{\prime}(r)\right], 
\label{12}
\end{equation}
where the prime denotes $r$ derivative. The temperature dependent bulk "mass" $\mu(r,T)$ of the boson field $M_N$ is related to that at zero temperature as follows:
\begin{equation}
\mu^{2}(r,T)=\frac{\mu^{2}}{f^{\frac{3}{5}}(r)}.
\label{13}
\end{equation}
 The five-dimensional mass  $\mu^{2}$ is expressed by means of the conformal dimension $\Delta =N+L$ of the interpolating operator dual to the  meson. $N$ is the number of partons and $L=max |L_z|$ is the quark orbital angular momentum.  $N=2$ for our $\rho$ meson and $L=0$ for the  meson ground state. For this meson the spin $J$ is $J=1$ and the expression for $\mu^{2}R^{2}$ obtains simple form \cite{25}:
\begin{equation}
\mu^{2}R^{2}=(\Delta-1)(\Delta-3).
\label{14}
\end{equation} 

  The axial gauge $M_{z}(x,r,T)=0$ is chosen for the vector field $M_N$, and the Klauza - Klein (KK) expansion is performed as follows:
\begin{equation}
M_{\mu}(x,r,T)=\sum_{n}{M_{\mu n}(x)\Phi_{n}(r,T)}.
\label{15}
\end{equation}
Here $M_{\mu n}(x)$ are KK modes wave functions corresponding to meson states, $\Phi_n (r,T)$ are their temperature-dependent profile functions,  and $n$ is the radial quantum number.
	
 Equation  of  motion  for  the  vector  field  will  be  reduced           
to the Schrödinger-type equation with the following replacement: $\phi_{n}(r,T)=e^{-\frac{B_{T}(r)}{2}}\Phi_n (r,T)$ with $B_{T}(r)=\varphi(r,T)-A(r)$. In the rest frame of the vector field the equation of motion (e.o.m.) will  give us the following equation for the  $\phi_{n}(r,T)$ profiles: 
\begin{equation}
\left[-\frac{d^{2}}{dr^{2}}+U(r,T)\right]\phi_{n}(r,T)=M_{n}^{2}(T)\phi_{n}(r,T).
\label{16}
\end{equation}
Here $U(r,T)$ is the effective potential and is the sum of the  temperature-dependent and nondependent parts:
\begin{equation}
U(r,T) =U(r) + \Delta U(r,T).
\label{17}
\end{equation}
Explicit forms of the $U(r)$ and $\Delta U(r,T)$ terms were given as
\begin{equation}
U(r)=k^{4}r^{2}+\frac{(4m^{2}-1)}{4r^{2}},
\label{18}
\end{equation}
\begin{equation}
\Delta U(r,T)=2\rho(T)k^{4}r^{2}.
\label{19}
\end{equation}
Here $m=N+L-2$ and, for the $\rho$ meson with two partons, it equals to $m=L$.
In the low temperature case the meson mass spectrum $ M_{n}^{2}$ is written as the following sum of zero and finite temperature parts:
\begin{equation}
 M_{n}^{2}(T) =\ M_{n}^{2}(0)+\Delta M_{n}^{2}(T),
 \label{20}
\end{equation}
\begin{equation}
\Delta M_{n}^{2}(T)=\rho(T)M_{n}^{2}(0) + \frac{R\pi^{4}T^{4}}{k^{2}},
\label{21}
\end{equation}
\begin{equation}
M_{n}^{2}(0)=4k^2\left(n+\frac{m+1}{2}\right), \\ R =(6n-1)(m+1).
\label{22}
\end{equation}
Finally, the solution of Eq. (\ref{16}) for the bulk profile $\phi_{n}(r,T)$ was given in the following form \cite{25}:
 \begin{equation}
 \phi_{n}(r,T)=\sqrt{\frac{2\Gamma(n+1)}{\Gamma(n+m+1)}}K^{m+1}r^{m+\frac{1}{2}}e^{-\frac{K^{2}r^{2}}{2}}L_{n}^{m}(K^{2}r^{2}).
 \label{23}
 \end{equation}
This solution coincides with the one at zero temperature case \cite{8} with the replacements $z\rightarrow r$, $K(T) \rightarrow k$ in it.
\section{BREAKING OF CHIRAL SYMMETRY AT FINITE TEMPERATURE}
The pseudo-scalar field $X$, which transforms under the bifundamental representation of $SU(2)_{L}\times SU(2)_{R}$, is introduced into the AdS/QCD models in order to perform breaking of this chiral symmetry group of the model using the Higgs mechanism in \cite{30,31,32,33}. The action for this field has the form
\begin{equation}
S_{X}=\int d^{4}xdr\sqrt{g}e^{-\varphi(r,T)}Tr\left[|DX|^{2}+3|X|^{2}\right].
\label{24}
\end{equation}
Here $D^{M}$ is the covariant derivative, which includes the minimal couplings with the $A^{M}_{L,R}$ gauge fields,
\begin{equation}
    D^{M}X = \partial^{M}X - iA^{M}_{L}X+XA^{M}_{R}=\partial^{M}X-i\left[M_M,X\right]-i\{A_M,X\}.
    \label{25}
\end{equation}
Since here we deal only with the vector field, we  ignore  the last term in Eq. (\ref{25}). The solution of the equation of motion for the $X$ field at zero temperature was widely described in  earlier works and we shall not repeat it here. We shall just recall the following vacuum expectation  value for this field, which was found in the Ref. \cite{30}:
\begin{equation}
\left< X \right>= \frac{1}{2}am_{q}z+\frac{1}{2a}\Sigma z^{3}=v(z).
\nonumber
\end{equation}
Here $m_{q}$ is the mass of $u$ and $d$ quarks, $\Sigma=<0|\bar{q}q|0> $ is the value of chiral condensate at zero temperature, and $a=\sqrt{N_{c}}/(2\pi)$. When the thermal case of this  solution was considered in Ref. \cite{25}, the temperature was taken into account by replacing  the  cold condensate with the thermal one and the $z$ coordinate with $r$:
\begin{equation}
\left< X(r,T)  \right> = \frac{1}{2}am_{q}r+\frac{1}{2a}\Sigma (T)r^{3}=v(r,T).
\label{26}
\end{equation}
In \cite{25, 26, 27} it was supposed that the temperature dependence of the $ \Sigma(T)=<0|\bar{q}q|0>_{T} $ quark condensate is identical to the temperature dependence of the dilaton parameter  $K^2(T)$:
\begin{equation}
K^2 (T)=k^2\frac{\Sigma(T)}{\Sigma}.
\label{27}
\end{equation}
In addition, it was conjectured that  the relation at zero temperature between the quark condensate  $\Sigma$  and the number of flavors  $N_f$,  the condensate parameter $B$,  and the  pseudo-scalar meson decay constant $F$ in the chiral limit
\begin{equation*}
\Sigma= - N_{f}BF^{2}
\end{equation*}
 holds for the finite temperature case as well:
\begin{equation}
\Sigma(T)= - N_{f}B(T)F^{2}(T).
\label{28}
\end{equation}
 Then according to Eqs. (\ref{7}) and (\ref{27}) we can write \cite{25}:
\begin{equation}
\Sigma(T)=\Sigma\left[1+\rho(T)\right].
\label{29}
\end{equation}
Let us note that the relation (\ref{29}) is valid until $T^6$ degrees of temperature. The $ F(T)$ and  $B(T)$ dependencies  were studied in \cite{25}.
\section{NUCLEON PROFILE FUNCTION AT FINITE TEMPERATURE}
In the AdS/QCD models we have to introduce two bulk fermion fields $\left(N_1, N_2\right)$ in order to describe two independent chiral components of nucleons \cite{31, 32} on the boundary. Let us present the solution of the equation of motion for the fermion fields describing boundary nucleons in this thermal soft-wall model. The action for the thermal fermion field $N(x,r,T)$ is written as \cite{26}:
\begin{equation}
S=\int d^{4}x dre^{-\varphi(r,T)}\sqrt{g}{\bar{N}}(x,r,T)D_{\pm }(r)N(x,r,T),
\label{30}
\end{equation}
where  $D_{\pm }(r)$ denotes the covariant derivative and has the explicit form 
\begin{equation}
D_{\pm}(r) =\frac{i}{2}\Gamma^{M}\left[\partial_{M}-\frac{1}{4}\omega_{M}^{ab}\ \left[\Gamma_{a}\Gamma_{b}\right]\right]\mp\left[\mu_F(r, T)+U_{F}(r, T)\right].
\label{31}
\end{equation}
 Here  $\mu_F (r, T)$ is the five dimensional mass of  thermal fermion field $N(x,r,T)$, and it is expressed in terms of $f(r, T)$ as follows:
\begin{equation}
\mu_F(r, T)=\mu_F\ f^{\frac{3}{10}}(r, T).
\label{32}
\end{equation}
Zero temperature mass $\mu_F$ is determined by the following equation between the number of partons ($N_{B}=3$) in the composite fermion and the orbital angular momentum $L$ ($L=0$ for the nucleons considererd here): 
\begin{equation}
\mu_F=\emph N_{B}+\emph L-\frac{3}{2},
\label{33}
\end{equation}
The temperature-dependent potensial $U_{F}(r,T)$ in Eq. (31) for the fermions is related to the zero temperature one as follows:
\begin{equation}
U_{F}(r, T)=\varphi (r,T)/f^{\frac{3}{10}}(r, T)
\label{34}
\end{equation}
 and the nonzero components of spin connection $ \omega_{M}^{ab} $ are given by
\begin{equation}
\omega_{M}^{ab}=(\delta_{M}^{a}\delta_{r}^{b}-\delta_{M}^{b}\delta_{r}^{a})\ r f^{\frac{1}{5}}(r, T). 
\label{35}
\end{equation}
The Dirac matrices $\Gamma^{A}$ in the $\left[\Gamma^{M},\Gamma^N\right]$ commutator in Eq. (\ref{31}) are related to those in the reference frame by the $\Gamma^{M} =e_{a}^{M}\Gamma^{a}$  relation, where $e_{a}^{M}=r\times diag \{\frac{1}{f(r)},1,1,1,-f(r)\} $ are the inverse vielbeins. The reference frame $\Gamma^{a}$ matrices are chosen as  $ \Gamma^{a}=(\gamma^{\mu },\ -i\gamma ^{5}) $. Using the axial gauge $N_{5}(x,r,T)=0$ we decompose the AdS fermion field into the left- and right-chirality components:
\begin{equation}
N(x,r,T)=N^{R}(x,r,T)+N^{L}(x,r,T),
\label{36}
\end{equation}
which are defined as usual ones
$ N^{R}(x,r,T)= \frac{1-\gamma^{5}}{2}N$,   $N^{L}(x,r,T)=\frac{1+\gamma^{5}}{2}N$
with properties
$\gamma^{5}N^{L}=-N^{L}$, $\gamma^{5}N^{R}=N^{R}$.
The Kaluza-Klein expansion for the $N^{L,R}$ chiral components will be written in terms of the temperature-dependent profile functions  $\Phi_{n}^{L,R}(r,T)$ as follows:
\begin{eqnarray}
N^{L,R}(x,r,T)=\sum_{n} N_{n}^{L,R}(x)\Phi_{n}^{L,R}(r,T).
\label{37}
\end{eqnarray}
For the nucleons the spin $\emph{J}=\frac{1}{2}$. For this case the $\Phi_{n}^{L,R}(r,T)$ profiles will be written with the prefactors 
\begin{equation}
\Phi_{n}^{L,R}(r,T)=e^{-\frac{3}{2}A(r)} F_{n}^{L,R}(r,T).
\label{38}   
\end{equation}
 After the substitution of these profile functions into the equations of motion in the rest frame of nucleon $(\vec{p}=0)$, we obtain the following form of the e.o.m. \cite{26} for the $F_{n}^{L,R}(r,T)$ profile functions:
\begin{equation} 
\left[\partial_{r}^2+U_{L,R}(r,T)\right]F_{n}^{L,R}(r,T)=M_{n}^{2}(T)F_{n}^{L,R}(r,T).
\label{39}
\end{equation}
The temperature-dependent spectrum $M_{n}^{2}(T) $ has a quantization similar to that in the zero temperature case, 
\begin{equation}
M_{n}^{2}(T)=4K^2(T)\left(n+m+\frac{1}{2}\right)=4k^2\left(1+\rho (T)\right)\left(n+m+\frac{1}{2}\right).
\label{40}
\end{equation}
$U(r,T)$ in Eq. (\ref{39}) is the effective potensial at finite temperature for the fermion field, and it can be decomposed into zero and finite temperature-dependent terms as follows: 
\begin{eqnarray}
U_{L,R}(r,T) =U_{L,R}(r)+\Delta U_{L,R}(r,T),
\nonumber \\
\Delta U_{L,R}\left(r,T\right) =2\rho (T)k^{2}\left(k^{2}r^{2}+m \mp \frac{1}{2}\right).
\label{41}
\end{eqnarray}
Here
\begin{equation}
m=N+L-\frac{3}{2}.
\label{42}
\end{equation}
Solutions to the Eq. (\ref{39}) are the following finite temperature profile functions for nucleons \cite{26}:
\begin{eqnarray}
F_{n}^{L}(r,T)=\sqrt{\frac{2\Gamma (n+1)}{\Gamma (n+m_{L}+1)}}K^{m_{L}+1}r^{m_{L}+\frac{1}{2}}e^{-\frac{K^{2}r^{2}}{2}}L_{n}^{m_{L}}\left(K^{2}r^{2}\right), \nonumber\\
F_{n}^{R}(r,T)=\sqrt{\frac{2\Gamma (n+1)}{\Gamma (n+m_{R}+1)}}K^{m_{R}+1}r^{m_{R}+\frac{1}{2}}e^{-\frac{K^{2}r^{2}}{2}}L_{n}^{m_{R}}\left(K^{2}r^{2}\right),
\label{43}
\end{eqnarray}
 where
$m_{L,R}=m\pm\frac{1}{2}$.
The profile functions $\Phi_{n}(r,T)$ and $F_{n}(r,T)$ obey normalization conditions 
\begin{equation}
\int_{0}^{\infty}dr e^{-\frac{3}{2}A(r)}\Phi_{m}^{L,R}(r,T)\Phi_{n}^{L,R}(r,T))=\int_{0}^{\infty }dr F_{m}^{L,R}(r,T) F_{n}^{L,R}(r,T)=\delta_{mn} 
\label{44} 
\end{equation} 
and coincide with those  for the zero temperature case with the replacements $r\rightarrow z$, $K(r,T) \rightarrow k$.    
\section{ BULK INTERACTION AND FINITE TEMPERATURE COUPLING CONSTANT}
 To   derive   the  $\rho$ meson-nucleon   thermal   coupling   constant in the AdS=CFT framework, we shall follow the
calculation procedure used in the zero temperature case in
Refs. \cite{30,31, 15, 8}. o this end we should construct a 
Lagrangian for the interaction in the bulk between the 
thermal vector and fermion bulk fields. Then, identifying 
the bulk partition function in the AdS-Schwarzschild 
background with the one in thermal QCD, we shall obtain
the expression for the thermal nucleon current interacting  with the thermal vector meson. For the finite temperature 
soft-wall model considered here the interaction action will 
be an integral of $L_{int.}$ multiplied by the exponent of the  thermal dilaton field. The five-dimensional interaction 
action in the bulk of AdS-Schwarzschild space-time will
be written as follows:
 \begin{equation}
 S_{int}=\int {d^{4}x}dr\sqrt{g}L_{int}e^{-\varphi(r,T)}.
 \label{45}
 \end{equation}
 According    to    the    holographic    principle,    the    generating
functional $Z_{AdS}$ of the bulk theory is identical to the 
generating functional $Z_{QCD}$ of the QCD theory on the 
UV boundary of this space-time:
 \begin{equation}
 Z_{AdS}=e^{iS_{int}} = Z_{QCD}.
 \label{46}
 \end{equation}
 So, in order to find the nucleon current (more precisely, a vacuum expectation value of the current), which interacts with the $\rho$ meson in boundary QCD theory, we can take the of the bulk  functional $Z_{AdS}$ from the boundary value of the bulk vector field $M_{\mu}^{a}(q)$ as follows:
 \begin{equation}
 <J_{\mu}^a>=-i\frac{\delta Z_{AdS}}{\delta M_{\mu}^{a}(q)}|_{M_{\mu}^{a}=0} .
 \label{47}
 \end{equation}
 There are several kinds of interactions between the spinor
 and vector bulk fields, and $L_{int}$ consists of terms describing
 these interactions. Since we have the gauge fields in the bulk,
 the first term is a minimal gauge interaction term of the vector
   field  with  the  current  of  fermions  in  the  bulk,

\begin{equation}
L_{MNN}^{(0)}(T)=\bar{N_{1}}e_{A}^{M}\Gamma^{A}M_{M}N_{1}+\bar{N_{2}}e_{A}^{M}\Gamma^{A}M_{M}N_{2}.
\label{48}
\end{equation} 
Next,      terms      are      connected      with      the      bulk      spinor           
field’s five-dimensional “magnetic moments,” which are 
described by $\Gamma^{MN}$. Four-dimensional components of this 
tensor correspond to the magnetic moments of fermions in  the reference frame. By means of these “moments” the 
terms of the Lagrangian can be constructed in the bulk 
of space-time, which have “Lorentz,” gauge, and parity 
invariant interactions with the vector field. The first such 
term is the following five-dimensional generalization of the 
usual four-dimensional magnetic interaction:
\begin{eqnarray}
L_{MNN}^{(1)}(T)=ik_{1}e_{A}^{M}e_{B}^{N}\left[\bar{N_{1}}\Gamma^{AB}(F_{L})_{MN}N_{1}-\bar{N_{2}}\Gamma^{AB}(F_{R})_{MN}N_{1}+h.c.\right] \nonumber \\
=ik_{1}e_{A}^{M}e_{B}^{N}\left[\bar{N_{1}}\Gamma^{AB}F_{MN}N_{1}-\bar{N_{2}}\Gamma^{AB}F_{MN}N_{1}+h.c.\right]+ axial \enspace vector \enspace term,
\label{49}
\end{eqnarray}
where $F_{MN}=\partial_MM_N-\partial_NM_M$ is  the field strength tensor of the $M_N$ vector field. 

The second such magnetic moment terms was constructed in \cite{32} and has the form
\begin{eqnarray}
L_{MNN}^{(2)}(T)=\frac{i}{2}k_{2}e_{A}^{M}e_{B}^{N}\left[\bar{N_{1}}X\Gamma^{AB}(F_{R})_{MN}N_{2}+\bar{N_{2}}X^{+}\Gamma^{AB}(F_{L})_{MN}N_{1}-h.c. \right] \nonumber \\ 
=\frac{i}{2}k_{2}e_{A}^{M}e_{B}^{N}\left[\bar{N_{1}}X\Gamma^{AB}F_{MN}N_{2}+\bar{N_{2}}X^{+}\Gamma^{AB}F_{MN}N_{1}+ axial \enspace vector \enspace term \right].
\label{50}
\end{eqnarray}
In addition to the magnetic moment interaction  this term includes with interaction with the $X$ field and as a  result of this interaction it changes the chirality of the fermion fields. As mentioned, the bulk scalar field $X$ was introduced as one, which changes the chirality of  the boundary nucleons and is
 expressed with the quark condensate  $\Sigma$in the boundary
 theory. Interpretation of this term is the interaction between
 the bulk fermions and the gauge fields by means of the
 magnetic moments of the fermions and an interaction with
 the $X$  field (the background field of the condensate). As a
 result of such a tripartite interaction, the chirality of the
 fermions is changed. In boundary QCD theory this term
 describes the nucleon- $\rho$ meson-quark condensate coupling with the change of chirality of the nucleons. The $k_{1}$ and $k_{2}$ constants were determined in the hard-wall model in the zero temperature case \cite{32}. Thus, the total "magnetic" -type Lagrangian is the following sum of these two terms:
\begin{equation}
L^{\prime}_{MNN}(T)=L_{MNN}^{(1)}(T)+L_{MNN}^{(2)}(T).
\label{51}
\end{equation}

Higher order bulk fields can be included in the
 Lagrangian terms; however, we neglect them in the
       approximation  here.

 Having explicit expressions of thermal profile functions
 of the bulk fields we can calculate the terms of thermal
 action in the momentum space and then take the variational
 derivative (\ref{47}) from these terms. This variation gives us
 the following contribution of each Lagrangian term to the
   nucleon  current:
 \begin{equation}
 \left<J_{\mu}\left(p^{\prime},p;T\right)\right>=g_{\rho NN}(T)\int dp^{\prime} dp \bar{u}(p^{\prime})\gamma_{\mu}u(p),
 \label{52}
 \end{equation}
where $g_{\rho NN}(T)$ is the integral over the holographic coordinate $r$ and it corresponds to the thermal minimal coupling constant according to holographic identification the bulk and boundary currents. 
Now we can write down the contribution of each Lagrangian term to the  $g_{\rho NN}(T)$ constant.

The contribution coming from the $\emph L_{MNN}^{(0)}$  Lagrangian is denoted by $g_{\rho NN}^{(0)nm}(T)$ and its integral expression is equal to the following one:

 \begin{eqnarray}
g_{\rho NN}^{(0)nm}(T)=\int_{0}^{\infty }\frac{dr}{r^{4}}e^{-K^{2}r^{2}}M_{0}(r,T)\left[F_{1L}^{*(n)}(r,T)F_{1L}^{(m)}(r,T)\right.\nonumber\\
\left.+F_{2L}^{*(n)}(r,T)F_{2L}^{(m)}(r,T)\right].
\label{53}     
\end{eqnarray}

Here we have used relations between the profile functions of the bulk fermion fields as $F_{1L}^{(s)} = F_{2R}^{(s)}$, $F_{1R}^{(s)}= -F_{2L}^{(s)}$, which are correct  for parity even states of the nucleons. . $M_{0}(r,T)$ is the profile function of a vector meson in the ground state.

In the Lagrangian expressions (\ref{49}) and  (\ref{50}) the $\Gamma^{MN}F_{MN}$ matrix is the sum of two kinds of terms-namely, $\Gamma^{5\nu}F_{5\nu}$ and  $\Gamma^{\mu\nu}F_{\mu\nu}$. Since these terms have different physical meanings, it is useful to present the contributions of these terms separately. In the  total Lagrangian $L^{\prime}_{MNN}(T)$  the $\Gamma^{5\nu}F_{5\nu}$ terms contribute to the $g_{\rho NN}$ constant, and  the contribution of this term is expressed as follows:
\begin{eqnarray}
g_{\rho NN}^{(1)nm}(T)=-2\int_{0}^{\infty }\frac{dr}{r^{3}}e^{-K^{2}r^{2}}M_{0}^{\prime}(r,T)\left[k_{1}\left(F_{1L}^{*(n)}(r,T)F_{1L}^{(m)}(r,T)\right.\right.\nonumber\\
\left.\left.-F_{2L}^{*(n)}(r,T)F_{2L}^{(m)}(r,T)\right)+k_{2}v(r,T)\left(F_{1L}^{*(n)}(r,T)F_{2L}^{(m)}(r,T)-F_{2L}^{*(n)}(r,T)F_{1L}^{(m)}(r,T)\right)\right].
\label{54}    
\end{eqnarray}

Here the prime on $M_{n}$ denotes the derivative over r. The  $\Gamma^{\mu\nu}F_{\mu\nu}$ term makes the following contrubution to the expression: 
\begin{eqnarray}
f_{\rho NN}^{nm}(T)=-4m_{N}\int_{0}^{\infty }\frac{dr}{r^{3}}e^{-K^{2}r^{2}}{M_{0}}(r,T)\left[k_{1}\left(F_{1L}^{*(n)}(r,T)F_{1R}^{(m)}(r,T)-F_{2L}^{*(n)}(r,T)F_{2R}^{(m)}(r,T)\right) \right.\nonumber\\
\left.+k_{2}v(r,T)\left(F_{1L}^{*(n)}(r,T)F_{2R}^{(m)}(r,T)-F_{2L}^{*(n)}(r,T)F_{1R}^{(m)}(r,T)\right)\right]. 
\label{55}   
\end{eqnarray}
where $m_{N}$ is the mass of the nucleon.  $f_{\rho NN}^{nm}(T)$  is interpreted as the contribution of the nucleon-$\rho$ meson interaction by means of the  magnetic moment of the nucleon.
The total coupling constant $g_{\rho NN}^{s.w.}(T)$ is the following sum of the previous coupling constants:
\begin{equation}
g_{\rho NN}^{s.w.}(T)=g_{\rho NN}^{(0)nm}(T)+g_{\rho NN}^{(1)nm}(T).
\label{56}
\end{equation}
The $g_{\rho NN}^{(0)nm}(T)$ coupling constant is interpreted as the "strong charge" of this interaction.
\section{Numerical analysis}
The numerical analysis of the $g_{\rho NN}^{(0)nm}(T)$
coupling constant consists of a numerical calculation of the integrals for the constants $g_{\rho NN}^{(0)nm}(T)$,
 $g_{\rho NN}^{(1)nm}(T)$, and $f_{\rho NN}^{nm}(T)$  and of numerically drawing their temperature dependencies by means of $Matematica$ package. We present our numerical results for the choice of parameters for the two flavor $N_{f}=2$ case with the pseudoscalar decay constant in the chiral limit $F=0.087$ $GeV$; for the three flavor  $N_{f}=3$ case with $F=0.1$ $GeV$; for the  four flavor $N_{f}=4$ case with $F=0.13$ $GeV$; and  for the five flavor $N_{f}=5$ case $F=0.14$ $GeV$. These sets of parameters were taken  from \cite{25}.
 The $k$ parameter was fixed at the value $k=0.383$ $GeV$ in \cite{25}. The parameter was fixed at the value We have free parameters  $k_{1}$ and $k_{2}$ were fixed at the values  $k_{1}=-0.78$ $ GeV^{3}$,  $k_{2}=0.5 $ $GeV^{3}$ in  \cite{32}. Here we do not consider these constants to be temperature-dependent ones, and we use these values in our numerical analysis. The $\Sigma=(0.368)^{3}$ $GeV^{3}$ and  $m_{q}=0.00145$ $GeV$ values of these parameters were found using  the fitting of the $\pi$ meson mass \cite{34}. 
To have an idea of relative contributions of different terms of Lagrangian, we present results for the temperature dependencies of the $g_{\rho NN}^{(0)nm}$, $g_{\rho NN}^{(1)nm}(T)$,  $g_{\rho NN}^{(s.w.)}(T)$, and $f^{nm}_{\rho}(T)$ coupling constants separately. In the figures below, the blue graph curve represents the $g_{\rho NN}^{(0)nm}$, the orange curve shows the $g_{\rho NN}^{(1)nm}(T)$, the green curve shows the $g_{\rho NN}^{(s.w.)}(T)$, and the red one shows the $f^{nm}_{\rho}(T)$ at  finite temperature. 
Finally, we have considered these dependencies for the first excited state $N(1440)$ of the nucleons as well. We repeat plotting graphs for the different numbers of flavors $N_f$ and $F$. We observe that all graphs in the Figs. 1-8 converge at one temperature value, which varies slightly in the different cases. Changing of the parameter values does not changes the picture in the figure and leads to only a slight  deformation of the  shape in the graphs. 
\section{discussion}
In this paper we have studied the temperature dependence of the strong coupling constant of the $\rho$ meson with
 the nucleons within the soft-wall model of AdS/QCD.
 We have plotted this dependence for each term in the
 coupling constant and have observed that all terms
 become zero at the same point near the Hawking
 temperature. (This point shifts slightly from case to
 case, which we think is related to the calculation
 accuracy.) The result here is reasonable from a physical
 interpretation point of view. Since the confinement-deconfinement phase transition occurs at the Hawking
 temperature and there are no hadrons after this temperature, we have obtained a zero value for the coupling
 constant between the hadrons below this temperature. If
we move from the inverse direction of the temperature 
axis, from higher temperatures to the Hawking temperature, we observe that the $g_{\rho NN}$ coupling constant at the 
point near the Hawking temperature becomes nonzero not sharply but smoothly. This may be interpreted as follows: 
strong interactions between the $\rho$ mesons and nucleons 
emerge not simultaneously with the hadronization, but 
instead just after the start of cooling of the formed hadron medium. This interpretation may be of use for the 
understanding processes at the early stages of the formation of the Universe. To have a complete physical 
picture of the finite temperature interactions between the 
hadrons, similar investigations are in process for the pion 
and axial vector–meson interactions with the nucleons in the framework of this formalism.

\newpage
\begin{figure}[!ht]
\centering
\includegraphics[scale=0.7]{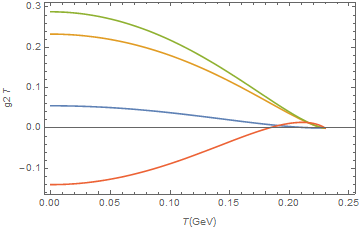}
\caption{Comparison of $g_{\rho NN}^{(0)nm}$, $g_{\rho NN}^{(1)nm}(T)$, and $f^{nm}_{\rho}(T)$ coupling constants at finite temperature for $N_{f}=2$, $F=0.087 MEV$.}
\label{fig:Figure1}
\end{figure}
\begin{figure}[!ht]
\centering
\includegraphics[scale=0.7]{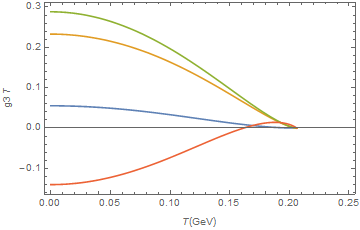}
\caption{Comparison of $g_{\rho NN}^{(0)nm}$,
$g_{\rho NN}^{(1)nm}(T)$, $g_{\rho NN}^{(s.w.)}(T)$ and $f^{nm}_{\rho}(T)$ coupling
constants at the parameter values $N_{f}=3$, $F=0.1 GEV$.}
\end{figure}
\begin{figure}[!ht]
\centering
\includegraphics[scale=0.7]{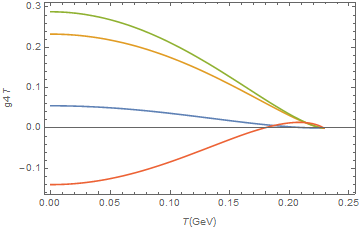}
\caption{Temperature dependence of the $g_{\rho NN}^{(0)nm}$,
$g_{\rho NN}^{(1)nm}(T)$, $g_{\rho NN}^{(s.w.)}(T)$ and $f^{nm}_{\rho}(T)$ coupling constants at the parameter values $N_{f}=4$, $F=0.13$ $GeV$.} 
\end{figure}
\begin{figure}[!ht]
\centering
\includegraphics[scale=0.7]{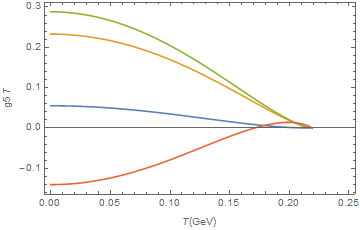}
\caption{The $g_{\rho NN}^{(0)nm}$,
$g_{\rho NN}^{(1)nm}(T)$, $g_{\rho NN}^{(s.w.)}(T)$ and $f^{nm}_{\rho}(T)$ coupling
constants at the parameter values $N_{f}=5$, $F=0.14$ $GeV$.}
\end{figure}
\begin{figure}[!ht]
\centering
\includegraphics[scale=0.7]{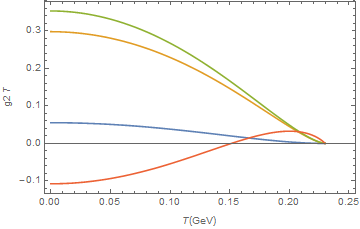}
\caption{The $g_{\rho NN}^{(0)nm}$,
	$g_{\rho NN}^{(1)nm}(T)$, $g_{\rho NN}^{(s.w.)}(T)$ and $f^{nm}_{\rho}(T)$ coupling
	constants for the first excited nucleons $N(1440)$ at the parameter values $N_{f}=2$, $F=0.087$ $GeV$.}
\label{fig:Figure 5}
\end{figure}
\begin{figure}[!ht]
\centering
\includegraphics[scale=0.7]{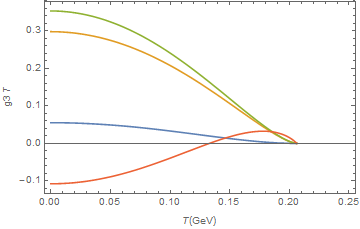}
\caption{The $g_{\rho NN}^{(0)nm}$,
	$g_{\rho NN}^{(1)nm}(T)$, $g_{\rho NN}^{(s.w.)}(T)$ and $f^{nm}_{\rho}(T)$ coupling
	constants for the first excited nucleons $N(1440)$ at the parameter values $N_{f}=3$, $F=0.1$ $GeV$.}
\end{figure}
\begin{figure}[!ht]
\centering
\includegraphics[scale=0.7]{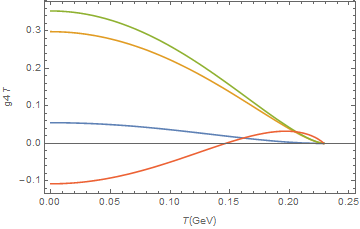}
\caption{The $g_{\rho NN}^{(0)nm}$,
	$g_{\rho NN}^{(1)nm}(T)$, $g_{\rho NN}^{(s.w.)}(T)$ and $f^{nm}_{\rho}(T)$ coupling
	constants for the first excited nucleons $N(1440)$ at the parameter values $N_{f}=4$, $F=0.13$ $GeV$.}
\end{figure}
\begin{figure}[!ht]
\centering
\includegraphics[scale=0.7]{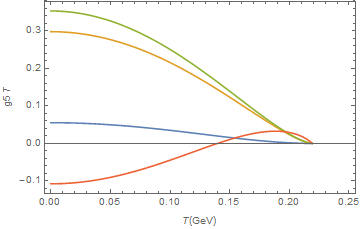}
\caption{The $g_{\rho NN}^{(0)nm}$,
	$g_{\rho NN}^{(1)nm}(T)$, $g_{\rho NN}^{(s.w.)}(T)$ and $f^{nm}_{\rho}(T)$ coupling
	constants for the first excited nucleons $N(1440)$ at the parameter values $N_{f}=5$, $F=0.14$ $GeV$.}
\end{figure}
\end{document}